\documentclass[a4paper,11pt,reqno,final]{amsart}

\usepackage{amsmath}
\usepackage{amssymb}
\usepackage{amsthm}
\usepackage{graphicx}
\usepackage{dcolumn}
\usepackage{bm}
\usepackage{psfrag}      
\usepackage{subfig}
\usepackage{pstricks}
\usepackage{setspace}  

\numberwithin{equation}{section}

\newtheorem{theorem}{Theorem}[section]
\newtheorem{lemma}[theorem]{Lemma}
\newtheorem{proposition}[theorem]{Proposition}

\newtheorem{ex}{\textsc{Example}}

\newcommand{\grad}{\mathop{\rm grad}}

\def\R{\mathbb{R}}

\def\K{\mathbb{K}}
\def\P{\mathbb{P}}

\def\rmd{{\mathrm d}}

\DeclareMathAlphabet{\bi}{OML}{cmm}{b}{it}
\DeclareMathAlphabet{\bcal}{OMS}{cmsy}{b}{n}

\begin{document}

\title[Behavioural and Dynamical Scenarios]{Behavioural and Dynamical Scenarios for Contingent Claims Valuation in Incomplete Markets}

\author[L. Boukas]{L. Boukas}
\address[L. Boukas]{University of the Aegean, Samos, Greece}
\email{lboukas@aegean.gr}
\author[D. Pinheiro]{D. Pinheiro}
\address[D. Pinheiro]{CEMAPRE, ISEG, Universidade T\'ecnica de Lisboa, Lisboa, Portugal}
\email{dpinheiro@iseg.utl.pt}
\author[A. A. Pinto]{A. A. Pinto}
\address[A. A. Pinto]{Departamento de Matem\'atica, Universidade do Minho, Braga, Portugal}
\email{aapinto@math.uminho.pt}
\author[S. Z. Xanthopoulos]{S. Z. Xanthopoulos}
\address[S. Z. Xanthopoulos]{University of the Aegean, Samos, Greece}
\email{sxantho@aegean.gr}
\author[A. N. Yannacopoulos]{A. N. Yannacopoulos}
\address[A. N. Yannacopoulos]{Athens University of Economics and Business, Athens, Greece}
\email{ayannaco@aueb.gr}
\date{}

\begin{abstract}
We study the problem of determination of asset prices in an incomplete market proposing three different but related scenarios. One scenario  uses a market game approach whereas the other two are based on risk sharing or regret minimizing considerations. Dynamical schemes modeling the convergence of the buyer's and of the seller's prices to a unique price are proposed.


\end{abstract}

\maketitle

\section{Introduction}
Realistic markets are incomplete, due to either the existence of frictions or to the nonexistence of the necessary assets needed to achieve the complete replication of any contingent claim by linear combinations of available (traded) assets. Incomplete markets is a very interesting field of economic theory and finance, which through seminal studies (see e.g. \cite{Geanakoplos_Polemarchakis}, \cite{Karatzas_Shreve} and \cite{Magill_Shafer}) has led to important results that have helped the community to reach a deeper understanding of the function of financial markets. 

In this work we study the problem of determination of prices in an incomplete market. While there is a rich literature on this field, the majority of these works focuses on the determination of bounds on the prices that are consistent with general equilibrium considerations. It is well known for instance, that in an incomplete market set up there is no longer a unique pricing kernel (martingale measure) and the existence of more than one pricing kernel may at best point out a whole band of prices that do not allow for arbitrage opportunities. The determination of a single price, at which an asset will eventually be traded in this market, out of this whole band, is still an open problem. There exists an extensive and very interesting literature on the subject, focusing on the determination of the upper and lower hedging prices (see e.g. \cite{Karatzas_Shreve}) as well as a number of suggestions on the price selected by the market (e.g. Kuhlback-Leibner or related entropy  minimization
criteria \cite{Hobson}, \cite{Staum}) which lead to interesting implications, some of which are testable with real market data, however, a complete theory of price selection in incomplete markets is still missing.
 
The aim of the present paper is to contribute to this literature providing  types of scenarios on price selection in incomplete markets.
The scenarios are  based on behavioural considerations and lead to interesting results. We present these scenarios within a 
one period model setting, so that the basic concepts and underlying ideas are made clear and technical details are kept to a minimum. Then the passage to a multiperiod model should follow without any major difficulty and we plan to present it in future work together with the case of the continuous model. 

The rest of the paper is organized as follows. In Section \ref{sec:DETERMINATION} we discuss the determination of  the upper and lower prices of a contingent claim in an incomplete market for risk averse agents via utility pricing. Although this is not a new issue (see e.g. \cite{Gamba_Pelizzari}), we include it here to fix ideas and notation.  In Section \ref{sec:3SCENARIOS} we address the problem of the determination of one commonly accepted price for the contingent claim with the use of three different scenarios. The first approach is a market game approach, the second is a risk sharing approach whereas the third approach is one in which the agents update their beliefs about the possible states of the world, in a way which is consistent with the minimization of total regret. In Section \ref{sec:DYNAMIC} we propose dynamic mechanisms that may describe the bargaining procedure between the agents, leading  to prices consistent with any of the above scenarios, and we study their stability.
Furthermore, we discuss the robustness of the previously mentioned dynamical mechanisms with respect to uncertainty. In Section 5 we summarize and conclude.

\section{Non-arbitrage upper and lower prices via utility pricing}\label{sec:DETERMINATION}
We consider a one period market model with $N$ tradable assets $a_1,...,a_N$ and $K$ states of the world $\omega_1,...,\omega_K$. One of the assets is a riskless asset with return rate $r$ and without loss of generality we may assume that this asset is $a_1$. Let ${\bf p}=(p_1,...,p_N)$ denote the price vector of the $N$ assets at time $t=0$, i.e. for each $i=1,...,N$ we denote by $p_i$ the price at time $t=0$ of the asset $a_i$. Let also ${\bf D}=(d_{ij})_{{\stackrel{i=1,...,K}{j=1,...,N}}}$ denote the $K\times N$ matrix of the assets payoffs at time $t=1$, that is, $d_{ij}$ denotes the value of the asset $a_j$ at time $t=1$ when the prevailing state of the world is $\omega_i$. In the sequel we may use the notation $d_j(\omega_i)$ instead of $d_{ij}$ and we may write $d_j$ to indicate the $j-$th column vector of ${\bf D}$ without these causing any confusion.  We also assume that the value of the riskless asset at time $t=1$ is equal to 1 no matter what the prevailing state of the world is, i.e. $d_1(\omega_i)=1$ for all $i=1,...,K$ (clearly, $p_1=1/(1+r)$). Moreover, without loss of generality, we assume  that none of the assets $a_1,...,a_N$ is redundant, that is, none of the columns of the payoff matrix ${\bf D}$ can be expressed as a linear combination of the remaining columns of the matrix. Furthermore we assume that $N<K$, that is, we face an incomplete markets situation where there are more states of the world than assets and thus the existing assets are not enough to reproduce every possible contingent claim at time $t=1$.

As it is known from the theory of mathematical finance, the incompleteness of the market leads to the existence of an infinity of risk neutral measures, all of which are consistent with the absence of arbitrage, that is, with general equilibrium theory.

 Let ${\mathcal M}$ be the family of risk neutral measures. This  family consists of $K$--tuples $Q=(q_1,...,q_K)$ such that $0<q_i<1$ for all $i=1,...,K$ and $q_1+...+q_K=1$, satisfying also the condition
\begin{eqnarray}
r=q_1 \left ( \frac{d_j(\omega_1)-p_j}{p_j} \right) + ... + q_K \left ( \frac{d_j(\omega_K)-p_j}{p_j} \right)
\nonumber
\end{eqnarray}
for all $j=1,...,N$. 

Any claim $F$ at $t=1$ that is contingent on the prevailing state will have price which is consistent with the absence of arbitrage assumption 
\begin{eqnarray}
P = \frac{1}{1+r}E_{Q}[F], \hspace{2mm} Q \in {\mathcal M} \ .
\nonumber
\end{eqnarray}
Therefore, each contingent claim may have an infinity of prices in a band, all of which are consistent with the assumption of  the absence of arbitrage  in the market. 

The question we wish to address in this paper is the following:
\begin{quote}
Out of this infinity of prices, which one will be actually realized?
\end{quote}

In  the absence of a unique risk neutral price for the asset, we resort to utility valuation. An expected utility maximizing risk averse agent, who operates under some probability measure $P$, chooses a reservation price of the contingent claim (which in fact is a non-arbitrage price) in such a way that she will be indifferent between either transacting or not transacting the contingent claim. In general, the reservation price of the seller of the contingent claim is different than that of the buyer of the contingent claim.  

\subsection{Allocation of wealth and indifference pricing}

Assume that an agent's preferences are described by a smooth utility function $U( \cdot )$ such that $U' >0$ (strict non-satiation) and  $U'' <0$ (strict risk aversion). These two properties guarantee the strict quasiconvexity of the utility function (see. e.g. \cite{MasCollel_Whinston_Whinston}).
   
The agent desires to allocate her initial wealth in the market in a way that her expected utility is maximized.
Suppose that the agent will allocate her initial wealth $W_0$ to the $N$ assets $a_1,...,a_N$ in proportions $\bm{\pi} =(\pi_1,...,\pi_N)$, i.e. proportion  $\pi_i$ of the initial wealth  will be placed in asset $a_i$ and $\pi_1+...+\pi_N=1$.  The final wealth of this portfolio at time $t=1$ is the random variable $W_{1}$ given by   
\begin{eqnarray}
W_{1}=\left[ W_0\sum_{j=1,...,N}\frac{\pi_j}{p_j}d_j(\omega_i)\right]_{i=1,...,K} \nonumber \ .
\end{eqnarray}
The agent will choose $\bm{\pi}$ so as to maximize $E[U(W_{1})]$ where the expectation is taken under a measure $Q=(q_1,...,q_K)$, reflecting the beliefs of the agent about the probabilities of occurrence of the future states of the world. So $q_i$ denotes the perceived by the agent probability that the state $\omega_i$ will occur.

It is well known that under the assumptions made here this problem has a solution. Let $\bm{\pi}_*$ denote the solution to this maximization problem and let $U_*$ denote the resulting maximized expected utility.
Clearly both ${\bm{\pi}_*}$ and $U_*$ are functions of the probability measure $Q$ and of the initial wealth $W_0$. 

Assume now that the seller has initial wealth $W^S_0$ and preferences described by a utility function $U^S(\cdot )$ and that she has issued a contingent claim with payoff to the buyer, at time $t=1$, given by the vector $d_{N+1}=\left(d_{N+1}(\omega_i)\right)_{i=1,...,K}$.
If the seller had not sold the contingent claim, she would have allocated proportions ${\bm{\pi}^S} =(\pi^S_1,...,\pi^S_N)$ of her wealth $W^S_0$ to the $N$ assets, according to the previous discussion, in order to achieve maximum expected utility $U^S_*(W_0^S;Q^S)$ under her perceived probability measure $Q^S$ about the future states of the world.   

Assume now that the seller decides to issue the contingent claim and receives the price $p^S_{N+1}$ at time $t=0$. The seller desires now to invest her initial wealth $W^S_0+p^S_{N+1}$ among the $N$ assets $a_1,...,a_N$ in proportions ${\bar{\bm{\pi}}}^S =(\bar{\pi}^S_1,...,\bar{\pi}^S_N)$, i.e. proportion  $\bar{\pi}^S_i$ of the initial wealth  will be placed in asset $a_i$ and $\bar{\pi}^S_1+...+\bar{\pi}^S_N=1$.  The final wealth of this portfolio at time $t=1$ is the random variable $\bar{W}^S_{1}$ given by   
\begin{eqnarray}
 \bar{W}^S_{1}=\left[ (W^S_0+p^S_{N+1})\sum_{j=1,...,N}\frac{\bar{\pi}^S_j}{p_j}d_j(\omega_i)-d_{N+1}(\omega_i)\right]_{i=1,...,K} \nonumber \ .
\end{eqnarray}
The seller will choose $\bar{\bm{\pi}}^S$ so as to maximize $E[U^S(\bar{W}^S_{1})]$, where the expectation is taken under the measure $Q^S=(q^S_1,...,q^S_K)$, reflecting the beliefs of the seller about the probabilities of occurrence of the future states of the world. So $q^S_i$ denotes the perceived by the seller probability that the state $\omega_i$ will occur

Let $\bar{\bm{\pi}}^S_*$ denote the solution to this maximization problem and let $\bar{U^S_*}$ denote the resulting maximized expected utility.
Clearly both $\bar{\bm{\pi}}^S_*$ and $\bar{U^S_*}$ are functions of the probability measure $Q^S$, of the initial wealth $W^S_0+p^S_{N+1}$ and of the random payoff of the liability undertaken by the writer of the contingent claim $d_{N+1}$.                                                                              
The seller of the contingent claim will choose a price for this contract so that these two decisions leave her indifferent, i.e. she will choose a price $p^{S}_{N+1}$ as the solution of the equation
\begin{eqnarray}
U^{S}_{*}(W^{S}_{0};Q^{S})=\bar{U}^{S}_{*}(W^{S}_{0}+p^{S}_{N+1},\, d_{N+1}\, ; Q^{S}) \nonumber \ .
\end{eqnarray}
The solution of this equation will provide the seller's indifference or reservation price of the contingent claim.
 \\                                                                                                                                                                                                                                                                                                                                                                                                              
\\                                                                                   
Similarly, assume  that the buyer has initial wealth $W^B_0$ and preferences described by a utility function $U^B(\cdot )$ and that she decided to buy the contingent claim with payoff to the buyer, at time $t=1$, given by the vector $d_{N+1}=\left(d_{N+1}(\omega_i)\right)_{i=1,...,K}$.
If the buyer had not bought the contingent claim, she would have allocated proportions $\bm{\pi}^B =(\pi^B_1,...,\pi^B_N)$ of her wealth $W^B_0$ to the $N$ assets, according to the previous discussion, in order to achieve maximum expected utility $U^B_*(W^B_0;Q^B)$ under her perceived probability measure $Q^B$ about the future states of the world.   

Assume now that the buyer bought the contingent claim for a price $p^B_{N+1}$ at time $t=0$. The buyer desires now to invest her initial wealth $W^B_0-p^B_{N+1}$ among the $N$ assets $a_1,...,a_N$ in proportions $\bar{\bm{\pi}}^B =(\bar{\pi}^B_1,...,\bar{\pi}^B_N)$, that is, proportion  $\bar{\pi}^B_i$ of this wealth  will be placed in asset $a_i$ and $\bar{\pi}^B_1+...+\bar{\pi}^B_N=1$.  The final wealth of this portfolio at time $t=1$ is the random variable $\bar{W}^B_{1}$ given by   
\begin{eqnarray}
 \bar{W}^B_{1}=\left[ (W^B_0-p^B_{N+1})\sum_{j=1,...,N}\frac{\bar{\pi}^B_j}{p_j}d_j(\omega_i)+d_{N+1}(\omega_i)\right]_{i=1,...,K} \nonumber \ .
\end{eqnarray}
The buyer will choose $\bar{\bm{\pi}}^B$ so as to maximize $E[U^B(\bar{W}^B_{1})]$, where the expectation is taken under the measure $Q^B=(q^B_1,...,q^B_K)$, reflecting the beliefs of the buyer about the probabilities of occurrence of the future states of the world. So $q^B_i$ denotes the perceived by the buyer probability that the state $\omega_i$ will occur.

Let $\bar{\bm{\pi}}^B_*$ denote the solution to this maximization problem and let $\bar{U}^B_*(W^B_0-p^B_{N+1} ,\,  d_{N+1} ;\, Q^B) $ denote the resulting maximized expected utility.
Clearly both $\bar{\bm{\pi}}^B_*$ and $\bar{U}^B_*$ are functions of the probability measure $Q^B$, of the initial wealth $W^B_0+p^B_{N+1}$ and of the contingent claim $d_{N+1}$.

Similarly, as in the case of the seller, the buyer will choose her price so that she will stay indifferent between buying or not buying the contingent claim contract. The price $p^B_{N+1}$ will be given as the solution of the algebraic equation $$U^B_*(W^B_0;Q^B) = \bar{U}^B_*(W^B_0-p^B_{N+1}, \,  d_{N+1} ;\, Q^B) \ .$$
Assuming that such a solution exists, it is called the indifference or reservation price for the buyer of the contingent claim contract. In general, when markets are incomplete, the reservation price of a contingent claim for the seller and the reservation price of the same contingent claim for the buyer do not coincide.

\section{ Three scenarios for price selection } \label{sec:3SCENARIOS}
From the above we see that in general the buyer and the seller of the contingent claim may end up with different valuations of the contingent claim.
The buyer has her own valuation of the contingent claim $P_B=p^{B}_{N+1}={\mathcal{P}}_{B}(Q^B)$, which is implicitly defined by the relation $U^B_* = {\bar{U}}^B_*$ for some probability measure $Q^B$. 
Similarly, the seller has her own valuation of the contingent claim $P_S=p^{S}_{N+1}={\mathcal{P}}_{S}(Q^S)$, which is implicitly defined by the relation $U^S_* = {\bar{U}}^S_*$ for some probability measure $Q^S$.
These valuations reflect the beliefs of the two agents about the future states of the world and their degrees of risk aversion. The two agents may be thought as not taking any risk at these valuations with regard to their expected utilities.
At this point both the seller and the buyer are unaware of each others valuation.

In this section we propose three different but interrelated scenarios under which the buyer and the seller will eventually agree on a common price for the contingent claim contract, at which the trade will take place. 
\\
\subsection{Market games approach} 
We have  assumed that the two agents will eventually reach a price agreement and we want to explore the basic scenarios under which this can be achieved. We also assume that the two agents are impatient. Impatience leads them to act (even momentarily) as if they are entering a one bid sealed-auction and within this framework we consider that their objective is minimization of maximum regret (according to Linhart \cite{Linhart} in the minimax regret case there exists a unique strategy, in contrast to the case of maximization of profit objective, which is a generalization of the linear equilibrium of Chatterjee-Samuelson \cite{Chatterjee_Samuelson}). 

Let $P_B$ denote the value of the contingent claim to the buyer and similarly, let $P_S$ denote the value of the contingent claim to the seller. We assume that the support of the buyer's prior as to the distribution of $P_B$ is $[\alpha ,\,\, \beta ]$, equal to the support of the seller's prior as to the distribution of $P_S$. Let $\tilde{P}_B={\mathcal{P}}_B(\tilde{Q}^B)$  be the intended bid price of the buyer and $\tilde{P}_S={\mathcal{P}}_S(\tilde{Q}^S)$ be the intended ask price of the seller. Trade occurs if and only if $\tilde{P}_B\geq \tilde{P}_S$. If trade occurs, the price is $P=\frac{\tilde{P}_B+\tilde{P}_S}{2}$. Thus the buyer's profit is $\Pi_B= P_{B}-\frac{\tilde{P}_B+\tilde{P}_S}{2}$ if $\tilde{P}_B\geq \tilde{P}_S$, otherwise it is zero. On the other hand the seller's profit is $\Pi_S= \frac{\tilde{P}_B+\tilde{P}_S}{2}-P_{S}$ if $\tilde{P}_B\geq \tilde{P}_S$, otherwise it is zero. Let $\Pi_B^*=\max_{\tilde{P}_B}\Pi_B$ be the maximum profit of the buyer (i.e. the best the buyer could have done, had he known the seller's ask) and let $R_B=\Pi_B^*-\Pi_B$ be the maximum regret of the buyer, and similarly for the seller. Following Linhart \cite{Linhart}, and considering  minimization of maximum regret, the two agents will make their offers:
\begin{equation*}
\tilde{P}_B=
\begin{cases}
P_B & \qquad \alpha \leq P_B\leq \alpha+\frac{1}{4}(\beta-\alpha) \\
\frac{2}{3}P_B+\frac{1}{12}(\beta-\alpha) & \qquad \alpha+\frac{1}{4}(\beta-\alpha)\leq P_B\leq \beta
\end{cases}
\end{equation*}
\begin{equation*}
\tilde{P}_S=
\begin{cases}
\frac{2}{3}P_S+\frac{1}{4}(\beta-\alpha) & \qquad \alpha \leq P_S\leq \alpha +\frac{3}{4} (\beta-\alpha) \\
P_S &\qquad \alpha +\frac{3}{4}(\beta-\alpha)\leq P_S\leq \beta
\end{cases}
\end{equation*}

If $P_B-P_S\geq\frac{\beta -\alpha}{4}$ then $\tilde{P}_B\geq\tilde{P}_S$ and therefore a trade is achieved at the price $P=\frac{\tilde{P}_B+\tilde{P}_S}{2}$ while at the same time both the buyer and the seller have minimized their maximum regret and they have concluded the deal in just one step. Of course they have both achieved a better price than their initial valuations since $P\leq P_B$ and $P\geq P_S$.

If however $P_B-P_S <\frac{\beta -\alpha}{4}$ then $\tilde{P}_B < \tilde{P}_S$ and the trade is not achieved at this first step. However, considering that $\tilde{P}_B$ and $\tilde{P}_S$ are one to one functions of $P_B$ and $P_S$, respectively, each agent can work out the initial valuation of her counterparty, that is, the buyer obtains knowledge of $P_S$ and the seller obtains knowledge of $P_B$. By doing so they will encounter one of the following two possibilities:
\begin{eqnarray*}
P_B\geq P_S
\label{eqn:SIMPLETRADE} \\
P_B< P_S
\label{eqn:BARGAINTRADE}
\end{eqnarray*}
In the first case, when $P_B\geq P_S$, they will split the difference between their initial valuations, and the resulting price will be $\frac{P_B+P_S}{2}$ and the trade will conclude. Again they will have both achieved a better price than their initial valuations since $\frac{P_B+P_S}{2}\leq P_B$ and $\frac{P_B+P_S}{2}\geq P_S$ and their maximum regret will be zero.
It is worth noting here that up to now the risk aversion characteristics of the two agents have not entered the bargaining procedure. Nevertheless, they have already been taken into account when the initial valuations $P_B$ and $P_S$ were first determined by the buyer and the seller, respectively.
In the second case, when $P_B< P_S$, no trade is possible unless the two agents either decide to undertake some risk or to update their beliefs. These two interesting scenarios will be dealt in the following sections.


\subsection{The risk sharing approach }\label{sec:RISKSHARING}
Another scenario for the price determination in an incomplete market is the scenario of optimal risk sharing. In this scenario we assume that the transaction of the contingent claim has to take place necessarily.  In this case the agents are obliged to take the risk associated with the contract and the best they can do  is to choose the price that guarantees that both the buyer and the seller undertake the minimum risk. This will yield a unique price for the contingent claim, under standard assumptions on the utility functions of the buyer and the seller.

We present a version of this scenario in this section.

Suppose that the seller and the buyer have beliefs about the future states of the world $Q^S$ and $Q^B$, respectively, so that the contingent claim indifference price of the buyer is lower than that of the seller. Then there will be no trade unless each of the agents decides to undertake some risk. For each agent we will quantify risk as the difference of the maximum expected utility of  the wealth while holding the long or short position in the contingent claim and the maximum expected utility of the wealth  without the contingent claim.

Let $U^A_{*}(W^A;Q^A)$ be  the maximum expected utility of an agent $A$ that has not undertaken any position on the contingent claim, where $W^A$ is the initial wealth of the agent and $Q^A$ is the probability measure reflecting the beliefs of the agent about the future states of the world. 
Let us now assume that the seller issues the contingent claim at an initial price $P_S$ while simultaneously adopting a position in the underlying market so that she maximizes her expected utility. 
In this case the maximum expected utility achieved is equal to 
$\bar{U}^{S}_{*}(W^S+P_S,d_{N+1} ; Q^{S})$.
If the seller decides to undertake risk $\epsilon_S$ then the price $P_S$ corresponding to this risk position will be given by the solution of the equation
\begin{eqnarray}
U^{S}_{*}(W^S; Q^{S})-\bar{U}^{S}_{*}(W^S+P_S,d_{N+1}; Q^{S})=\epsilon_S \ .
\nonumber
\end{eqnarray}
We will denote the solution of this algebraic equation by $P_S(\epsilon_S)$. This depends on the risk undertaken as well as on the beliefs on the future states of the world. Note that $P_S(0)$ is the indifference price for the seller.

Let us now assume that the buyer buys the contingent claim at an initial price $P_B$ while simultaneously adopting a position in the underlying market so that she maximizes her expected utility.
The maximum utility is then
$\bar{U}^{B}_{*}(W^B-P_B,d_{N+1};Q^{B})$.
If the buyer decides to undertake risk $\epsilon_B$ then the price $P_B$ corresponding to this risk position will be given by the solution of the equation
\begin{eqnarray}
U^{B}_{*}(W^B ; Q^{B})-\bar{U}^{B}_{*}(W^B-P_B , d_{N+1} ; Q^{B})=\epsilon_B
\nonumber
\end{eqnarray}
We will denote the solution of this algebraic equation by $P_B(\epsilon_B)$. This also depends on the risk undertaken as well as on the beliefs on the future states of the world. Note that $P_B(0)$ is the indifference price for the buyer. 

In the next lemma we summarize some properties of the functions $P_S(\epsilon_{S})$ and $P_{B}(\epsilon_{B})$.
\begin{lemma}\label{thm:PRICEPROPERTIES}
Assume that the seller and the buyer make their decisions with expected utility functions $U^{S}(W)=E[u_{S}(W)]$ and $U^{B}(W)=E[u_{B}(W)]$, respectively. Furthermore, assume that $u_{i}{'}>0$ and $u_{i}''<0$ for $i=S,B$.\\
Then,\\
(i) $P_{S}$ is strictly decreasing  and strictly quasiconcave in $\epsilon_{S}$.\\
(ii) $P_{B}$ is strictly increasing and strictly quasiconvex in $\epsilon_{B}$.\\
(iii) The function $P_{S}(\epsilon_{S})-P_{B}(\epsilon_{B})$ is strictly quasiconcave.
\end{lemma} 
\begin{proof} 
$(i)$ According to the results of \cite{Gamba_Pelizzari}, under our assumptions for the utility function,  the portfolio optimization problem has a unique solution and the maximum expected utility  of the seller $\bar{U}^{S}_{*}(w , d_{N+1} ; Q^{S})$ is a strictly increasing function of its first argument $w$. According to the definition of $P_{S}(\epsilon_{S})$ we see that 
\begin{eqnarray}
U^{S}_{*}(w;Q^{S})-\bar{U}^{S}_{*}(w+P_S(\epsilon_{S}), d_{N+1} ; Q^{S}) = \epsilon_{S}
\label{eqn:PSDEFN}
\end{eqnarray}
 Take two values of risk undertaken $\epsilon_{S} > \epsilon_{S}'$. Using the definition of $P_{S}(\epsilon_{S})$ and $P_{S}(\epsilon_{S}')$, respectively, we obtain that
\begin{eqnarray}
U^{S}_{*}(w+P_{S}(\epsilon_{S}),d_{N+1};Q^{S})-U^{S}_{*}(w+P_{S}(\epsilon_{S}'),d_{N+1};Q^{S})=-(\epsilon_{S}-\epsilon_{S}') < 0
\nonumber
\end{eqnarray}
Since $U^{S}_{s}(w,d_{N+1};Q^{S})$ is strictly increasing in $w$ the above inequality guarantees that $P_{S}(\epsilon_{S}) <P_{S}(\epsilon_{S}')$, so that $P_{S}$ is strictly decreasing. 

We now show that $P_{S}(\epsilon_{S})$ is strictly quasiconcave. Taking the convex combination of (\ref{eqn:PSDEFN}) at $\epsilon_{S}$ and $\epsilon_{S}'$, respectively, we find that
\begin{eqnarray}
\lambda \bar{U}^{S}_{*}(w+P_S(\epsilon_S))+ (1-\lambda) \bar{U}^{S}_{*}(w+P_S(\epsilon_S'))=\bar{U}^{S}_{*}(w+P_S(\lambda \epsilon_{S}+(1-\lambda)\epsilon_{S}'))
\label{eqn:UTI}
\end{eqnarray}
where to make the notation simpler we have suppressed the dependence of the indirect utility function to the liability and the measure and kept explicitly only the dependence on the initial wealth.

Suppose now that $P_{S}(\epsilon_S) > P_{S}(\epsilon_{S}')$. Then, by the strict monotonicity
of the indirect utility function $\bar{U}^{S}_{*}(w)$ we obtain that $\bar{U}^{S}_{*}(w+P_S(\epsilon_{S}))>\bar{U}^{S}_{*}(w+P_S(\epsilon_{S}'))$ so that
\begin{eqnarray}
\bar{U}^{S}_{*}(w+P_S(\epsilon_{S}')) &<& \lambda \bar{U}^{S}_{*}(w+P_S(\epsilon_{S})) + (1-\lambda)  \bar{U}^{S}_{*}(w+P_S(\epsilon_{S}')) \nonumber \\
& < & \bar{U}^{S}_{*}(w+P_S(\epsilon_{S})) 
\label{eqn:UTI1}
\end{eqnarray}
Using equation (\ref{eqn:UTI}) we see that (\ref{eqn:UTI1}) implies that
\begin{eqnarray}
\bar{U}^{S}_{*}(w+P_S(\epsilon_{S}'))<\bar{U}^{S}_{*}(w+P_S(\lambda \epsilon_{S}+(1-\lambda)\epsilon_{S}')) < \bar{U}^{S}_{*}(w+P_S(\epsilon_{S}))
\end{eqnarray}
so that by the monotonicity of the indirect utility function we observe that $P_S(\lambda \epsilon_{S}+(1-\lambda)\epsilon_{S}')> P_S(\epsilon_{S}')$. Therefore $P_S(\lambda \epsilon_{S}+(1-\lambda)\epsilon_{S}')> \min (P_S(\epsilon_{S}),P_S(\epsilon_{S}'))$, which implies the strict quasiconcavity of $P_{S}(\epsilon_{S})$.

The proof of $(ii)$ is similar taking into account that the position of the buyer is the opposite than that of the seller. Finally $(iii)$ follows easily from $(i)$ and $(ii)$. 
\end{proof}

We may define the total risk undertaken by both agents by the convex combination
\begin{eqnarray}
R(\epsilon_{S},\epsilon_{B}):=\lambda \epsilon_{S}+(1-\lambda) \epsilon_{B} \ .
\nonumber
\end{eqnarray}
If $\lambda \in (0,1)$ this corresponds to sharing the total risk undertaken by the agents in the proportion $\lambda/(1-\lambda)$.

We then define the price of the contingent claim by the solution of the following optimization problem
\begin{eqnarray}
\min_{(\epsilon_S,\epsilon_B)} \lambda \epsilon_{S}+(1-\lambda) \epsilon_{B}
\nonumber \\
\mbox{subject to the constraint}
\label{eqn:PRIMAL} \\
P_S(\epsilon_S) \le P_B(\epsilon_B)
\nonumber
\end{eqnarray}
If this problem has a unique solution $(\epsilon_S^{*},\epsilon_B^{*})$ this would  lead to a unique price $P_S(\epsilon_S^{*})=P_B(\epsilon_B^{*})$. The solution of this problem is formally equivalent to the solution of the expenditure minimization problem and thus the prices $P_S(\epsilon_S^{*})$ and $P_{B}(\epsilon_{B}^{*})$ present an interesting analogy to the determination of the Hicksian demand function. 

As we shall see, a closely related problem to the risk sharing problem is the following ``dual'' problem 
\begin{eqnarray}
\max_{\epsilon_S,\epsilon_B} P_{B}(\epsilon_{B})-P_{S}(\epsilon_{S})
\nonumber \\
\mbox{subject to the constraint} \label{eqn:DUAL} \\
\lambda \epsilon_{S}+ (1-\lambda) \epsilon_{B} \le w
\nonumber 
\end{eqnarray}
The solution of the dual problem (\ref{eqn:DUAL}) deals with the problem of maximizing the stated price difference between the buyer and the seller under the constraint of undertaking a prespecified total amount of risk  $w$. The maximization of the price difference ensures the possibility of successful trading of the asset. The properties of the functions $P_{B}(\epsilon_{B})$ and $P_{S}(\epsilon_{S})$ will guarantee that the maximum is obtained at $P_{B}(\epsilon_{B})=P_{S}(\epsilon_{S})$.

The following proposition shows that these optimization problems admit unique solutions and sheds some light to their properties.

\begin{proposition}\label{thm:PRIMAL}
Under the conditions of Lemma \ref{thm:PRICEPROPERTIES} the following statements hold for the optimization problem (\ref{eqn:PRIMAL}):\\
(i) Problem (\ref{eqn:PRIMAL}) has a unique solution which will be denoted as $(\epsilon_{B}^{*}(\lambda),\epsilon_{S}^{*}(\lambda))$.\\
(ii) The function $\epsilon_{B}^{*}(\lambda)$ is nonincreasing in $\lambda \in (0,1)$ whereas the function $\epsilon_{S}^{*}(\lambda)$ is nondecreasing in $\lambda \in (0,1)$.\\
(iii) On the optimal risk bearing allocation $P_S(\epsilon_S)=P_B(\epsilon_B)$.
\end{proposition}
\begin{proof}
The optimization problem (\ref{eqn:PRIMAL}) is formally equivalent to the expenditure minimization problem in consumer theory and inherits most of its properties (see e.g. \cite{MasCollel_Whinston_Whinston}). The existence of the solution comes simply from the nonemptyness  of the constraint set $$B=\{ (\epsilon_B,\epsilon_S) \in {\R}^{2}_{+} \mid P_B(\epsilon_B)- P_S(\epsilon_S)\ge 0 \} \ .$$
This follows by continuity of $P_B(\epsilon_B)$ and $P_S(\epsilon_S)$.

The uniqueness of the solution comes from the strict convexity of the ``utility function'' $U(\epsilon_B,\epsilon_S)= P_B(\epsilon_B)- P_S(\epsilon_S)$, which is guaranteed by Lemma \ref{thm:PRICEPROPERTIES}.

We may further observe other properties of the optimal risk sharing allocation. The optimal risk undertaken by the seller $\epsilon_{S}^{*}(\lambda)$ is a nondecreasing function of $\lambda$ whereas the optimal risk undertaken by the buyer $\epsilon_{B}^{*}(\lambda)$ is a nonincreasing function of $\lambda$. This comes out easily from the continuity and monotonicity properties of the inverse of the utility functions we are using.

Finally, the local non-satiation property of the ``utility function'' $U(\epsilon_B,\epsilon_S)$ guarantees that at the optimal risk sharing allocation $P_{S}(\epsilon_{S})=P_{B}(\epsilon_B)$.
\end{proof}

We may also see through standard duality arguments that the solutions of problems (\ref{eqn:PRIMAL}) and (\ref{eqn:DUAL}) are related in the following sense:
 \begin{proposition}\label{thm:DUALITY}
(i) Given the risk sharing rule $\lambda$, let $(\epsilon_{B}^{*},\epsilon_{S}^{*})$ be a solution of problem (\ref{eqn:PRIMAL}) such that $\Delta P =P_{B}(\epsilon_{B}^{*})-P_{S}(\epsilon_{S}^{*})\ge 0$. Define $R=\lambda \epsilon_{B}^{*} + (1-\lambda) \epsilon_{S}^{*}$. If $R > \min \lambda \epsilon_{B} + (1-\lambda) \epsilon_{S}$ then $(\epsilon_{B}^{*},\epsilon_{S}^{*})$ is a solution to problem (\ref{eqn:DUAL}) given $(\lambda, R)$. For this solution of problem (\ref{eqn:DUAL}) we obtain that $\max\{ P_{B}(\epsilon_{B})-P_{S}(\epsilon_{S})\}=0$.\\
(ii) If $(\bar{\epsilon}_{B},\bar{\epsilon}_{S})$ is a solution to problem (\ref{eqn:DUAL}) given $(\lambda,R)$ then $(\bar{\epsilon}_{B},\bar{\epsilon}_{S})$ solves problem (\ref{eqn:PRIMAL}) subject to the constraint $P_{B}(\epsilon_{B})-P_{S}(\epsilon_{S}) \ge P_{B}(\bar{\epsilon}_{B})-P_{S}(\bar{\epsilon}_{S})$.
\end{proposition}

For the particular problem at hand, for which the price functions are differentiable, the calculation of the optimal risk beared will be given by the first order conditions
\begin{eqnarray}
&&\lambda \ge -\mu \frac{\rmd P_{S}}{\rmd \epsilon_{S}}
\nonumber \\
&&(1-\lambda) \ge \mu  \frac{\rmd P_{B}}{\rmd \epsilon_{B}}
\nonumber \\
&&\epsilon_{S}\left(\lambda +\mu \frac{\rmd P_{S}}{\rmd\epsilon_{S}}\right) + \epsilon_{B}\left((1-\lambda) -\mu \frac{\rmd P_{B}}{\rmd\epsilon_{B}}\right) =0
\nonumber
\end{eqnarray}
 for some $\mu \ge 0$.

\subsection{ Optimal choice of the agents market price of risk}\label{sec:OPTBELIEF}
In this class of scenarios we assume that each of the agents has an idea about the possible states of the world $Q^B$ and $Q^S$, respectively, but their beliefs are not ``rigid'' in the sense that whey are willing to change their attitudes  by observing the other agents. This is in contrast with the situation we described in the previous scenario, where each agent had a rigid belief about the possible value of the states of the world. We assume that each agent will choose the utility indifference price for the asset using as probability measure for the definition of the expected utility the ``personalized'' measure that corresponds to her own beliefs.  Therefore, assuming her beliefs are correct, she will not incur any losses (defined in the sense of utility of final wealth).
However, for generic choice of $Q^{B}$ and $Q^{S}$, the utility prices of the buyer and the seller will not coincide, thus leading to impossibility of trade. We assume that the agents, for some reasons, are determined to trade the contingent claim. Therefore, the agents review their attitudes towards the states of the world and adopt new beliefs $Q^{S}$, $Q^{B}$ so that their new indifference prices will become equal. Since the agents are flexible about their beliefs we assume that they fully trust their new attitudes towards states of the world and the new indifference prices incur no risk for either of the two agents.

Summarizing:
\begin{itemize}
\item Each agent starts with her beliefs $Q^{B}$ and $Q^{S}$ about the states of the world 
\item They state their indifference prices $P_{S}(Q^{S})$ and $P_{B}(Q^{B})$ under their beliefs about the states of the world. These prices incur no risk (with regard to expected utility)
\item In general $P_{S}(Q^{S}) \ne P_{B}(Q^{B})$
\item The agents update their beliefs about states of the world and state their new indifference prices
\item The process continues until the agents adopt beliefs such that   $P_{S}(Q^{S}) = P_{B}(Q^{B})$
\end{itemize}

The above procedure will in principle lead to a price such that $P_{S}(Q^{S}) = P_{B}(Q^{B})$. However, this price is still {\bf any} price within the band of prices defined by the upper and lower hedging prices. In what follows we propose some scenarios that may lead to a {\bf unique} price for the asset within this band.

Consider now the following argument: the seller of the asset will of course wish to obtain at least the price that corresponds to her beliefs about the future states of the world $Q^{S}=\overline{Q}$. By changing her belief to some new $Q^{S}$ the seller compromises to give away the potential ``extra'' profit corresponding to $\overline{Q}-Q^{S}$. 
On the other hand, the buyer of the asset would wish to obtain at most the price that corresponds to her beliefs about the future states of the world $Q^{B}=\underline{Q}$. By changing her belief to some new $Q^{B}$ the buyer compromises to give away the potential ``extra'' profit corresponding to $Q^{B}-\underline{Q}$.
The quantity $\lambda (\overline{Q}-Q^{S}) + (1-\lambda) (Q^{B}-\underline{Q})$ can be considered as the total potential loss of the two agents, where the parameter $\lambda$ gives us information on the way that this ``loss'' is divided between the two agents. In the case $\lambda=1/2$ the loss is equally divided between the agents. In the case $\lambda=0$ the buyer suffers all the potential loss, whereas in the case $\lambda=1$ the seller suffers all the potential loss. We have to stress here that this loss is the potential loss in the sense that it is the loss that the agents will suffer if they are unable to persuade the others to adopt their beliefs. In practice, since the agents are willing to fully adopt their new beliefs as the correct beliefs about the world and choose their prices by utility indifference pricing according to the new beliefs, there is no actual loss. That is why we call the loss, ``potential loss''.

We may consider now the case where the agents update their beliefs so that the total potential loss is minimized. This would correspond to choosing $Q^{S}$, $Q^{B}$ so that
\begin{eqnarray}
\min_{Q^{S}, Q^{B}}  \lambda d_{S}(\overline{Q}-Q^{S}) + (1-\lambda) d_{B} (Q^{B}-\underline{Q})
\nonumber \\
\mbox{ subject to the constraint }
\label{eqn:PRIMAL1} \\
P_S(Q^S) \le P_B(Q^B)
\nonumber
\end{eqnarray}
which would in turn lead to a price for the asset. In the above problem, $d_{B}$ and $d_{S}$ are distance functions in the unit simplex $\Delta^{K}$ quantifying the buyer's and seller's regret, respectively.

It is again interesting to consider the dual problem of (\ref{eqn:PRIMAL1}). This can be written as
\begin{eqnarray}
\max_{Q^{S},Q^{B}} P_{B}(Q^{B})-P_{S}(Q^{S})
\nonumber \\
\mbox{subject to the constraint}
\label{eqn:DUAL1} \\
\lambda d_{S}(\overline{Q}-Q^{S}) + (1-\lambda) d_{B}(Q^{B}-\underline{Q}) \le w
\nonumber
\end{eqnarray}

Problem (\ref{eqn:DUAL1}) corresponds to the maximization of the difference of the prices stated by the buyer and by the seller (so that trade is ensured) under the constraint of undertaking less than a fixed amount of total regret. The solution of problems (\ref{eqn:DUAL1}) and (\ref{eqn:PRIMAL1}) are related.

We assume that the beliefs $Q^{S}$ of the seller corresponds to some $\hat{Q}^{S} \in {\mathcal M}$ such that $P_{S}(Q^{S})=E_{\hat{Q}^{S}}[F^{*}]=:\hat{P}_{S}(\hat{Q}^{S})$, and similarly for the buyer. Within this context we may rewrite problems (\ref{eqn:PRIMAL1}) and (\ref{eqn:DUAL1}) in terms of the (linear) price functions $\hat{P}_{S}(\hat{Q}^{S})$ and $\hat{P}_{B}(\hat{Q}^{B})$ and the risk neutral measures $\hat{Q}^{S}$ and $\hat{Q}^{B}$. One may argue that within this context, the distance from a reference risk neutral measure may make more sense as regret.
This is  since the measures $\hat{Q} \in {\mathcal M}$ have more economic meaning, in terms of the marginal utility of consumption in the different states of the world,  than the measures $Q$ quantifying the beliefs of the agents. 

The following hold for both $P(Q)$ and $\hat{P}(\hat{Q})$.

\begin{lemma}\label{thm:PRICEPROP}		
(i) The price $P_{B}(Q^{B})$ is a continuous function of $Q^{B}\in \Delta^{K}$ and achieves a minimum value $\underline{P}_{B}$ and a maximum value $\overline{P}_{B}$.\\
(ii) The price $P_{S}(Q^{S})$ is a continuous function of $Q^{S} \in \Delta^{K}$ and achieves a minimum value $\underline{P}_{S}$ and a maximum value $\overline{P}_{S}$. \\
(iii) If $[\underline{P}_{B},\overline{P}_{B}]\cap [\underline{P}_{S},\overline{P}_{S}] \ne \emptyset$ then there exist $Q^{B},Q^{S} \in \Delta^{K}$ such that $P_{B}(Q^{B})\ge P_{S}(Q^{S})$.
\end{lemma}
\begin{proof}
 (i) The maximum theorem of Berge (see e.g. \cite{Berge}) guarantees the continuity of the function
 \begin{eqnarray}
 \bar{U}^{B}_{*}(W_{0}^{B}-P_B,d_{N+1};Q^{B})=\sup_{\theta} \sum_{i=1}^{K}q^{B}_{i} U^{B}(W_{0}^{B}-P_{B}+d_{N+1}(\omega_{i})+\theta \cdot \Delta G(\omega_i)),
 \nonumber
 \end{eqnarray} 
 with respect to the parameters $Q^{B}=(q^{B}_1,..., q^{B}_{K}) \in \Delta^{K}$. In the above, by 
$$\Delta G(\omega_{i})=(\Delta G_{1}(\omega_{i}),..., \Delta G_{N}(\omega_{i})) \ , i=1,..., K \ ,$$
 we denote the gains from the assets in the state of the world $\omega_{i}$.  The price $P_{B}(Q^{B})$ for the case where the buyer adopts the ``beliefs'' $Q^{B}$, is obtained as the solution of the equation
 \begin{eqnarray}
\bar{U}^{B}_{*}(W_{0}^{B}-P_{B}(Q^{B}),d_{N+1};Q^{B})=U_{*}^{B}(W_{0}^{B};Q^{B})
 \nonumber
 \end{eqnarray}
 By the continuity properties of the inverse function, we obtain continuity of $P_{B}(Q^{B})$ with respect to $Q^{B}\in\Delta^{K}$. Since $\Delta^{K}$ is a compact set, by Weierstrass theorem we obtain that $P_{B}(Q^{B})$ achieves a maximum and a minimum value in $\Delta^K$.\\
(ii) Follows along similar lines as $(i)$.\\
(iii) Follows by continuity of the functions $P_{B}(Q^{B})$ and $P_S(Q^{S})$.
\end{proof}

Using the above lemma, we may prove the existence of a price that minimizes the total regret of the buyer and the seller. 

\begin{proposition} Assume further that the functions $P_{B}(Q^{B})$ and $P_{S}(Q^{S})$ are strictly convex. Then, 
the following statements hold:\\
(i) There exists a unique solution to the minimization problem (\ref{eqn:PRIMAL1}).\\
(ii) The unique choice of the agents beliefs $Q^{S}$ and $Q^{B}$ corresponds to a unique price $P_{S}(Q^{S}) = P_{B}(Q^{B})$.\\
(iii) The maximization problem (\ref{eqn:DUAL1}) has an unique solution.
\end{proposition}

The assumption of strict convexity is quite natural for certain classes of utility functions. For instance, if the agents make their decisions using exponential utility functions (not necessarily with the same risk aversion coefficient) then detailed calculations show that the functions $P_{B}(Q^{B})$ and $P_{S}(Q^{S})$ are expressed in terms of  combinations of linear and logarithmic functions which may be shown to be strictly convex.

   The regret minimizing scenario may be generalized as follows. Consider that the agents work entirely in terms of the stated prices so that they update their prices directly rather than through their beliefs about the states of the world. The regret of the buyer and the seller may then be expressed as their divergence from the minimum and maximum price, respectively. Therefore, the total regret of the buyer and seller may be expressed as the convex combination 
   \begin{eqnarray}
   R=\lambda \phi_{B}(P_{B}-\underline{P}_{B}) + (1-\lambda) \phi_{S}(\overline{P}_{S}-P_{S}) \ ,
   \nonumber
   \end{eqnarray}
   where $\phi_{B}(x)$ and $\phi_{S}(x)$ are functions modeling the regret of the buyer and the seller, respectively, as a function of the deviation of the stated price from the preferred prices which are $\underline{P}_{B}$ for the buyer and $\overline{P}_{S}$ for the seller. These functions are considered as increasing functions of their argument. One possible choice for these functions may be the linear function, however, one may give these regret functions the same properties as utility functions, that is, they may be considered as increasing concave functions. 

The problem of agreeing to a common price for the contingent claim may then be expressed as the optimization problem
\begin{eqnarray}
&&\min_{(P_{B},P_{S})\in {\mathcal A}} \lambda \phi_{B}(P_{B}-\underline{P}_{B})+(1-\lambda) \phi_{S}(\overline{P}_{S}-P_{S})
\nonumber \\
&&\mbox{subject to}
\nonumber \\
&&P_{B}-P_{S} \ge 0
\nonumber
\end{eqnarray}
where ${\mathcal A}=[\underline{P}_{B},\overline{P}_{B}]\times [\underline{P}_{S},\overline{P}_{S}]$.
Assuming strict convexity of the regret function we may guarantee that the above problem leads to a unique common price.

\begin{proposition}
Assume that $\phi_{B}$ and $\phi_{S}$ are strictly convex increasing functions. Then,
the regret minimization problem has a unique solution.
\end{proposition} 

\section{Dynamic mechanisms for price selection and robustness with respect to uncertainty}\label{sec:DYNAMIC}
In the previous section we have proposed certain scenarios that lead to the choice of a price for an asset in an incomplete market. However, the above scenarios give a characterization of the static problem of choosing the final price of the asset. In this section we address the problem of the dynamics that will lead the agents to the adoption of the price in which the asset will finally be traded, which may be the one proposed by either of the above types of scenarios. Therefore, in this section we propose dynamic mechanisms by which the trading will take place. 

For the sake of concreteness and brevity  
let us only consider dynamic mechanisms in the context of scenarios of the type presented in section \ref{sec:RISKSHARING}. The same arguments may be easily generalized in the study of dynamic mechanisms for other types of scenarios e.g. those of section \ref{sec:OPTBELIEF}.

Consider the following situation: The agents will have to complete their transactions on the contingent claim by period $0$ of the model. At time $0^{-}$ the agents enter in a trading 'game' which may consist of several periods, the objective of which is that the agents update their attitudes towards risk so that they arrive at a single price. 
We will denote by $\epsilon_{B}(n)$ and $\epsilon_{S}(n)$ the amount of risk undertaken by the buyer and the seller of the contingent claim, respectively, at the $n$ trading period.

The risk updating scheme is given by a dynamical system of the form
\begin{eqnarray}
\epsilon_{S}(n+1)&=&\epsilon_{S}(n) + f_{1}(\epsilon_{S}(n),\epsilon_{B}(n)) (P_{S}(\epsilon_{S})-P_{B}(\epsilon_{B}))
\nonumber \\
\epsilon_{B}(n+1)&=&\epsilon_{B}(n) + f_{2}(\epsilon_{S}(n),\epsilon_{B}(n)) (P_{S}(\epsilon_{S})-P_{B}(\epsilon_{B})) \ ,
\nonumber
\end{eqnarray}
where $f_{1}(\epsilon_{S}(n),\epsilon_{B}(n))$ and $f_{2}(\epsilon_{S}(n),\epsilon_{B}(n))$ correspond to the velocity of update of the risk undertaken by each agent and will be specified later on. Their exact forms correspond to behavioural assumptions on the way that the agents update the risk they are willing to undertake, by observing the amount of risk undertaken by the other agent as induced by the prices stated.

To simplify the analysis further, we assume that the time scales at which the bargaining is made at period $0^{-}$ are very short, and at any rate negligible compared to the time period between $T=0$ and $T=1$. This implies that we may approximate the above dynamical system by a differential equation of the form
\begin{eqnarray}
\frac{\rmd \epsilon_{S}}{\rmd t}&=& f_{1}(\epsilon_{S}(t),\epsilon_{B}(t)) (P_{S}(\epsilon_{S}(t))-P_{B}(\epsilon_{B}(t)))
\nonumber \\
\frac{\rmd \epsilon_{B}}{\rmd t}&=& f_{2}(\epsilon_{S}(t),\epsilon_{B}(t)) (P_{S}(\epsilon_{S}(t))-P_{B}(\epsilon_{B}(t)))
\nonumber 
\end{eqnarray}
where now by $\epsilon_{B}(t)$ and $\epsilon_{S}(t)$ we denote the amount of risk undertaken by the agents at the trading period $t$. 

By proper choice of the functions $f_1$ and $f_2$  we may guarantee the convergence of the above dynamical trading scheme to a single price for the asset. The following proposition provides conditions under which this convergence is guaranteed.

\begin{proposition}\label{thm:CONV}
Assume that the inequality
\begin{eqnarray}
\lambda := \frac{\partial P_{S}}{\partial \epsilon_{S}} f_1 + 
\frac{\partial P_{B}}{\partial \epsilon_{B}} f_2 \le -\epsilon  
\label{eqn:CONDCONV}
\end{eqnarray}
holds for some positive $\epsilon$. \\
Then, the solution of the dynamical trading scheme converges to $P_{S}(\epsilon_{S})=P_{B}(\epsilon_{B})$. 
\end{proposition} 
\begin{proof}
It is straightforward to observe that
\begin{eqnarray}
\frac{\rmd}{\rmd t}(P_{S}-P_{B})=\left( \frac{\partial P_{S}}{\partial \epsilon_{S}} f_1 + 
\frac{\partial P_{B}}{\partial \epsilon_{B}} f_2 \right ) (P_{S}-P_{B})
\nonumber \ .
\end{eqnarray}
If $x=P_{S}-P_{B}$ we see that $x$ satisfies the differential inequality
\begin{eqnarray}
\frac{\rmd x}{\rmd t} \le  -\epsilon x
\nonumber
\end{eqnarray}
which, by the Gronwall inequality, guarantees that $P_{S} \rightarrow P_{B}$ as $t\rightarrow \infty$. 
\end{proof}

A large number of risk updating schemes satisfy this condition. Different dynamic mechanisms may be obtained depending on behavioural assumptions. One example might be the steepest descend algorithm, guaranteeing the fastest descent to the isopricing manifold $P_S(\epsilon_S)=P_B(\epsilon_B)$. 

\begin{ex}
Assume that
\begin{equation*}
f_{1}=-\frac{\partial P_{S}}{\partial \epsilon_{S}} \ , \qquad f_{2}=\frac{\partial P_{B}}{\partial \epsilon_{B}}  \ .
\end{equation*}
This corresponds to a risk undertaking scheme according to which the seller undertakes risk proportionally to the sensitivity of her price with respect to changes in $\epsilon_{S}$, and proportionally to the difference between the stated prices of seller and buyer, while the buyer undertakes risk with a similar scheme. 
Then gradually the seller price lowers while the buyer price increases until these two prices meet at $P_{S}(\epsilon_{S}) = P_{B}(\epsilon_{B})$.

The properties of the price function $P_{S}(\epsilon_{S})$ and $P_{B}(\epsilon_{B})$ (see Lemma \ref{thm:PRICEPROPERTIES}) show that condition (\ref{eqn:CONDCONV}) holds, 
so that according to Proposition \ref{thm:CONV} the convergence to the single price $P_{S}(\epsilon_{S}) = P_{B}(\epsilon_{B})$ is guaranteed. 
\end{ex}

The convergence dynamics to the isopricing manifold (i.e. to one price for the buyer and the seller that coincide)  may be explicitly studied in certain cases of interest, e.g. in the case where the buyer's and the seller's prices are chosen via indifference pricing using exponential utility functions (see e.g. \cite{Xanthopoulos_Yannacopoulos}).

The above risk updating scheme guarantees that buyer and seller arrive at a single price but does not necessarily guarantee that this is the optimal risk sharing price. This can be done through a dynamical scheme that may approximate the solution of the optimal risk sharing price. To this end we may approximate the original optimization problem (\ref{eqn:PRIMAL}) 
with an unconstrained problem, with the use of an entropic barrier method (see e.g. \cite{Boyd_Vandenberghe}). According to this method we may approximate problem (\ref{eqn:PRIMAL}) by the unconstrained problem
\begin{eqnarray}
\min_{\epsilon_{B},\epsilon_{S}} \lambda \epsilon_{B}+(1-\lambda) \epsilon_{S} -\frac{1}{h} \log( P_{B}(\epsilon_{B})-P_{S}(\epsilon_{S}))
\nonumber
\end{eqnarray}
for arbitrary $h>0$, which is related to the order of approximation of the original problem. As is easily seen, the objective function becomes infinite when the constraint $P_{B}(\epsilon_{B}) \ge P_{S}(\epsilon_{S})$ is not met. A gradient method may then be applied for the solution of this problem, which may be treated as a risk updating scheme. For instance, assuming for ease of notation continuous time updating, we may consider the dynamical scheme
\begin{eqnarray}
\frac{\rmd\epsilon_{B}}{\rmd t}&=&s \left( \lambda  - \frac{1}{P_{B}-P_{S}}\frac{\partial P_{B}}{\partial \epsilon_{B}}\right) 
\nonumber  \\
\frac{\rmd\epsilon_{S}}{\rmd t}&=&s \left( (1-\lambda)  + \frac{1}{P_{B}-P_{S}}\frac{\partial P_{S}}{\partial \epsilon_{S}}\right) 
\nonumber
\end{eqnarray}
for some $s >0$.
This scheme will drive us to a common price for buyer and seller with the minimum aggregate risk. To see that, take first the case where $P_{B} < P_{S}$, i.e. the case where no trade is possible. Then $P_{B}-P_{S}<0$ and since $\frac{\partial P_{B}}{\partial \epsilon_{B}}>0$ we see that $\frac{\rmd\epsilon_{B}}{\rmd t}>0$ so the buyer will undertake more risk. This will raise the buyer's price. Similarly, since $\frac{\partial P_{S}}{\partial \epsilon_{S}}<0$ we see that $\frac{\rmd\epsilon_{S}}{\rmd t}>0$ so the seller will also undertake more risk and this will lower the seller's price. As a result the two agents will finally tend towards $P_{B} \simeq P_{S}$. In the case where $P_{B} > P_{S}$,  we may easily observe that the ``forces'' will reverse and will be such that both agent start undertaking less risk until the minimum aggregate risk is obtained.
This simplistic hand waving argument may be turned into a rigorous mathematical proof of convergence. What is important, is to observe that such arguments allow us to provide further behavioural rules according to which a unique price for a contingent claim in an incomplete market may be obtained.

The above dynamical schemes may be generalized using the notion of projected dynamical systems, widely used in the theory of optimization. We only introduce the concept here with a simple example and intend to return to this issue in the near future. We see by the properties of the risk sharing solution (see Proposition \ref{thm:PRIMAL}) that the risk sharing price will be obtained by the solution of the fixed point equation $P_S(\epsilon_{S})-P_B(\epsilon_{B})=0$ under the constraint $(\epsilon_{S},\epsilon_{B}) \in {\K}$ where ${\K} = \{(\epsilon_{S},\epsilon_{B}) \, \mid \, \lambda \epsilon_{S}+ (1-\lambda) \epsilon_{B} \le w, \, \lambda \in [0,1]\}$. In other words, it is the solution of a fixed point scheme constrained to take values in a closed convex set. This brings us to the situation of a number of problems studied by Flam (see e.g. \cite{Flam}). According to his results, the solution to the problem may be written in the equivalent form
\begin{eqnarray}
x = {\P}_{{\K}}[x+(P_{S}(x)-P_{B}(x))] \ ,
\nonumber
\end{eqnarray}
where $x=(\epsilon_{S},\epsilon_{B})$ and ${\P}_{{\K}}$ is the orthogonal projection onto the closed convex set $\K$. The introduction of the projection operator ensures that the constraint is satisfied. The formulation given above for the fixed point may be turned into a dynamical system scheme for the determination of such fixed point.
In discrete time, this dynamical scheme may be written as
\begin{eqnarray}
x_{n+1}={\P}_{{\K}}[x_{n}+s_{n} (P_{S}(x_{n})-P_{B}(x_{n}))] \ ,
\nonumber
\end{eqnarray}
where $s_{n}$ is a sequence of properly selected step sizes. Under proper choices of the step sizes sequence $\{s_{n}\}$ the dynamical scheme will converge to the equilibrium $P_{S}(x)=P_{B}(x)$. In a continuous time approximation this scheme becomes 
\begin{eqnarray}
\frac{\rmd x}{\rmd t}=\Lambda \{{\P}_{{\K}}[x-\alpha (P_{S}(x)-P_{B}(x))]-x\}
\nonumber
\end{eqnarray}
for $\Lambda$ and $\alpha$ positive constants. This projected dynamical system can be shown to be globally asymptotically stable under certain conditions. 
 
 Another choice could be the projected gradient dynamical system
 \begin{eqnarray}
\frac{\rmd x}{\rmd t}=\Lambda \{{\P}_{{\K}}[x-\alpha \grad R(x)]-x\} \ ,
\nonumber
\end{eqnarray}
where $R(x)$ is the regret function we wish to minimize.
Let us for example consider the projected gradient dynamical system that may drive the system to the minimum total regret in the case where we study the problem of regret minimization in the space of prices rather than in the space of beliefs or risks. In this case, $x=(P_1,P_2)=(P_S,P_B)$, and the convex subset ${\K}=\{(P_1,P_2) \in {\R}^{2} \mid P_{1} \ge P_{2} \}$. The projection operator onto this set is the linear operator
\begin{eqnarray}
\pi_{{\K}} = \frac{1}{2} \left (
\begin{array}{cccc}
1 & 1
\nonumber \\
1 & 1
\end{array}
\right ) {\bf 1}_{\{P_1<P_2\}} + I {\bf 1}_{\{P_1 \ge P_2\}} \ .
\nonumber
\end{eqnarray}
The projected gradient system then becomes
\begin{eqnarray}
\frac{\rmd P_1}{\rmd t}&=&-\frac{\alpha}{2}(P_1-P_2) + \frac{\alpha}{2} (R_{1}'-R_{2}')
\nonumber \\
\frac{\rmd P_2}{\rmd t}&=&\frac{\alpha}{2}(P_1-P_2) + \frac{\alpha}{2} (R_{1}'-R_{2}')
\nonumber
\end{eqnarray}
when $P_{1}<P_{2}$ and 
\begin{eqnarray}
\frac{\rmd P_1}{\rmd t}&=&- \frac{\alpha}{2} R_{1}'
\nonumber \\
\frac{\rmd P_2}{\rmd t}&=& \frac{\alpha}{2} R_{2}'
\nonumber
\end{eqnarray}
when $P_{1}\ge P_{2}$, for some positive constant $\alpha$.
This projected gradient system can be shown to converge exponentially fast to the minimal regret, using Lyapunov function techniques similar to those used in Proposition \ref{thm:CONV} and the properties of the projection operator onto convex sets. We refrain from giving a rigorous proof, which could follow along the lines of \cite{Antipin}. 

However, it is interesting to look at it from the behavioural point of view. The system of prices updates requires not only knowledge of the stated prices of the two agents, which of course are readily accessible in the bargaining process, but also some knowledge of the preferences of the two agents (which is modeled by the gradient of the regret function). One may easily argue that agent 1 may know her regret function $R_1$ but not necessarily the regret function $R_2$ of the other agent. In this respect,  schemes like the above, even though they guarantee convergence to an optimal price may not be appropriate, unless treated in a fashion that includes uncertainty about the preferences of the other agent.    

An obvious question that comes to mind is the robustness of the above mechanisms with respect to uncertainty as to the strategy for price update followed. In order to provide preliminary results to this question we will assume that the above mentioned dynamical mechanisms are perturbed by the presence of noise terms that model the possible uncertainty with respect to the strategy followed by the buyer and the seller. For the sake of simplicity and to keep the technicalities to the minimum possible, we will assume that uncertainty may be modeled by the introduction of Brownian motion terms. 

We will show that some of the mechanisms proposed may be robust in the presence of noise. 

In the presence of noise, the dynamical scheme for risk selection takes the form
\begin{eqnarray}
\rmd\epsilon_{S}(t)&=&f_{1} (P_{S}-P_{B})\rmd t + \sigma_1 (P_{S}-P_{B}) \rmd W(t)
\nonumber \\
\rmd\epsilon_{B}(t)&=&f_{2}(P_{S}-P_{B})\rmd t + \sigma_2 (P_{S}-P_{B}) \rmd W(t) \ ,
\nonumber
\end{eqnarray}
where $\rmd W(t)$ are the increments of a standard Brownian motion which, without loss of generality, is taken to be one dimensional.

The basic concern of this section is the following question: 
\begin{quote}
Under which conditions the above stochastic risk updating mechanism converges to a common price for the buyer and the seller of the asset?
\end{quote}

The following proposition offers an answer to this question.

\begin{proposition}\label{thm:CONVSTOCH}
Let
\begin{eqnarray}
R_1&:=&\frac{\partial P_{S}}{\partial \epsilon_{S}} f_1 -
\frac{\partial P_{B}}{\partial \epsilon_{B}} f_2 
\nonumber \\
R_2&=&\frac{1}{2} \left \{    \frac{\partial^2 P_{S}}{\partial \epsilon_{S}^2} \sigma_1^2 -
\frac{\partial^2 P_{B}}{\partial \epsilon_{B}^2} \sigma_2^2  \right\}
\nonumber \\
R_3&:=&\left\{ \frac{\partial P_{S}}{\partial \epsilon_{S}} \sigma_1 -
\frac{\partial P_{B}}{\partial \epsilon_{B}} \sigma_2 \right \}
\nonumber
\end{eqnarray}
Suppose that $R_2 \le 0$ and that there exist $K,\rho,\Sigma >0$ such that $\mid R_1 \mid \le K$ as well as that $\underline{\Sigma} \le R_3^{2}  \le \overline{\Sigma}$ for all $\epsilon_{S}$, $\epsilon_{B}$.

Then, if $\rho > K +\frac{\Sigma}{2}$  the trivial solution $P_{S}=P_{B}$ is almost surely exponentially stable, i.e. almost all sample paths of the solution will tend to the equilibrium $P_{S}=P_{B}$ exponentially fast.
\end{proposition}
\begin{proof}
Using It\^o's formula we find that
\begin{eqnarray}
\rmd(P_{S}-P_{B})&=&\left\{ \frac{\partial P_{S}}{\partial \epsilon_{S}} f_1 -
\frac{\partial P_{B}}{\partial \epsilon_{B}} f_2 \right \} (P_{S}-P_{B}) \rmd t
\nonumber \\
&+& \frac{1}{2} \left \{    \frac{\partial^2 P_{S}}{\partial \epsilon_{S}^2} \sigma_1^2 -
\frac{\partial^2 P_{B}}{\partial \epsilon_{B}^2} \sigma_2^2  \right\}(P_{S}-P_{B})^2 \rmd t \nonumber  \\
&+& \left\{ \frac{\partial P_{S}}{\partial \epsilon_{S}} \sigma_1 -
\frac{\partial P_{B}}{\partial \epsilon_{B}} \sigma_2 \right \} (P_{S}-P_{B}) \rmd W(t)
\nonumber
\end{eqnarray}
or if we define $x=P_{S}-P_{B}$ the stochastic evolution for the difference of the seller's and buyer's price satisfies
\begin{eqnarray}
\rmd x(t) = f(t,x(t)) \rmd t + g(t, x(t)) \rmd W(t) \ ,
\label{eqn:DIF}
\end{eqnarray}
where 
\begin{eqnarray}
&&f(t,x(t)):= R_1 x(t) + R_2 x(t)^2
\nonumber \\
&&g(t,x(t)):= R_3 x(t)
\nonumber 
\end{eqnarray}
with $R_1$, $R_2$ and $R_3$ defined as in the statement of the Proposition.

Consider the Lyapunov function $V(x,t)=\mid x \mid^2$. Then the following hold for  the action of the infinitesimal generator $L$ of the diffusion process (\ref{eqn:DIF}) on $V$
\begin{eqnarray}
&&LV(x,t) = 2 (R_1 x + R_2 x^2) x  + R_3^2 x^2 \le 2 R_1 x^2 + R_3^2 x^2 \le  (2 K + \overline{\Sigma})  x^2
\nonumber \\
&&\mid V_{x}(x,t)g(t,x) \mid^2 = 4 R_3^2 x^2  \ge 4 \underline{\Sigma}  x^4
\nonumber
\end{eqnarray}
Then, applying  a theorem on stochastic stability (see Theorem 3.3, Section 4.3  p. 121 in  \cite{Mao}), we see that
\begin{eqnarray}
\lim \sup_{t\rightarrow \infty} \frac{1}{t} \ln \mid x (t) \mid \le - \left ( \underline{\Sigma} - K -\frac{\overline{\Sigma}}{2}\right)
\nonumber
\end{eqnarray}
which guarantees  that the trivial solution $x=0$ is almost surely exponentially stable if $\underline{\Sigma } > K +\frac{\overline{\Sigma}}{2}$. 
\end{proof}

\begin{ex}
Consider the stochastic risk selection  scheme
\begin{eqnarray}
\rmd\epsilon_{S}&=&-f_1 (P_{S}-P_{B}) + \sigma_1 (P_{S}-P_{B}) \rmd W(t)
\nonumber \\
\rmd\epsilon_{B}&=&f_2 (P_{S}-P_{B}) + \sigma_2 (P_{S}-P_{B}) \rmd W(t)
\nonumber
\end{eqnarray}
where $f_1 >0$, $f_2 >0$.

The properties of the price functions $P_{S}(\epsilon_{S})$ and $P_{B}(\epsilon_{B})$ guarantee that the conditions of Proposition \ref{thm:CONVSTOCH} hold so that convergence to the single price $P_{S}(\epsilon_{S})=P_{B}(\epsilon_{B})$ is obtained. 
\end{ex}

Similar results may be obtained for the other dynamical schemes proposed above in the presence of noise. Many of these schemes may be turned into stochastic schemes and their long time behaviour can be studied using results from stochastic stability theory (see e.g. \cite{Flam}). We intend to provide further results on this problem in future work.

\section{Conclusion}
In this work we address the question of price formation in incomplete markets. We elaborate within the framework of a one period discrete model
so that the concepts and the ideas behind price formation will be highlighted and the technical details will be kept to a minimum. Then the passage to a multiperiod model should be straightforward and we plan to present it in future work together with the case of the continuous model. 

It is well known that in an incomplete markets setting, if equivalent martingale measures exist, they are not unique. Therefore, this leads to more than one possible price, all of which consistent with the absence of arbitrage arguments.  Other criteria will therefore be needed in order to select the price at which a particular asset is traded in an incomplete market.
Several  criteria have been proposed in the literature for the selection of the measure chosen to price a particular asset in an incomplete market, the majority of which is based, to the best of our knowledge, on the minimization of entropy related functions. Such functions quantify the ``distance'' between the true statistical measure of the market and the equivalent martingale measure chosen by the agents in the market.

In this paper we have chosen to take an alternative route and propose three different, but ultimately related, scenarios for the price selection in incomplete markets. All scenarios assume that the participating agents have some initial beliefs about the distribution of the future states of the world. Based on these beliefs, each of them has in mind an initial non-arbitrage valuation of the contingent claim, according to which no risk is assumed utility-wise. However, these initial valuations do not coincide in general.  

The first scenario is a market game where the buyer and the seller bargain on the price of the contingent claim and choose the bargaining strategy that minimizes maximum regret. Given their initial valuations, this mechanism offers a unique bargaining strategy, that will lead to at most one unique price (depending on their initial valuations). 

The second scenario which leads to a unique price for the asset is based on the concept of risk sharing. In this scenario we assume that each of the agents has firm beliefs about the future prices of the world, but deliberately undertakes some risk so that the transaction will be made possible. The unique price of the asset is defined by the solution of the optimization problem, in which the risk undertaken by each agent is chosen so that  a convex combination of the  risks undertaken by the agents is minimized, under the constraint that the transaction is made possible, i.e. under the constraint that the buyer's price is greater or equal than the seller's price.  

The third scenario models the situation where the two agents do not have firm beliefs about the future states of the world but they are willing to update their beliefs as part of the bargaining procedure. Their quoted prices thus do not entail any risk but there is some potential loss, to which we call regret in this work. The potential loss for agent 1 comes about from not being able to persuade  agent 2 to accept her original beliefs (that would lead to the best possible price for her) and similarly for agent 2. 

In the third scenario a unique price is chosen by the solution of the optimization problem in which the beliefs are chosen so that the convex combination of the regrets of the two agents is minimized under the constraint that the transaction eventually takes place.  

These three scenarios offer plausible ways in which prices may be chosen in an incomplete market. This first part of our analysis deals with the static problem of specifying which price is finally selected.  

The second part of the paper deals with the proposal of dynamic mechanisms that will lead the agents to the price at which the contingent claim is traded. This section is more general in the sense that we study general dynamic mechanisms that will lead to a common price between buyer and seller, which may or may not be the one proposed in the three scenarios of part one.
These dynamical mechanisms are reminiscent of the Walras tatt\^onement scenario in general equilibrium considerations and add to the general literature on how markets are led to their ``equilibrium'' states. In this section arguments from the theory of dynamical systems are used to provide general schemes that will lead to a common price for the asset. The stability of these mechanisms is studied as well as their robustness with respect to random perturbations.

\section*{Acknowledgments}

We thank the Calouste Gulbenkian Foundation, PRODYN-ESF, POCTI, and POSI by FCT and Minist\'erio da Ci\^encia, Tecnologia e Ensino Superior, Centro de Matem\'atica da Universidade do Minho, CEMAPRE, and Centro de Matem\'atica da Universidade do Porto for their financial support. 

S. Xanthopoulos would like to acknowledge that this project is co-funded by the European Social Fund and National Resources - (EPEAEK-II) PYTHAGORAS.

D. Pinheiro would also like to acknowledge the financial support from ``Programa Gulbenkian de Est\'imulo \`a Investiga\c{c}\~ao 2006'' and FCT - Funda\c{c}\~ao para a Ci\^encia e Tecnologia grant with reference SFRH / BPD / 27151 / 2006. \\

\bibliography{BPPYX09}

\begin{thebibliography}{10}

\bibitem{Antipin}
A.~S. Antipin.
\newblock Minimization of convex functions on convex sets by means of
  differential equations.
\newblock {\em Differential Equations}, 30:1365--1375, 1994.

\bibitem{Berge}
C.~Berge.
\newblock {\em Topological Spaces}.
\newblock Dover (New York), 1997.

\bibitem{Boyd_Vandenberghe}
S.~Boyd and L.~Vandenberghe.
\newblock {\em Convex Optimization}.
\newblock Cambridge University Press (New York), 2004.

\bibitem{Chatterjee_Samuelson}
K.~Chatterjee and W.~Samuelson.
\newblock Bargaining under incomplete information.
\newblock {\em Operations Research}, 31:835--851, 1983.

\bibitem{Flam}
S.~D. Flam.
\newblock Approaches to economic equilibrium.
\newblock {\em Journal of Economic Dynamics and Control}, 20:1505--1522, 1996.

\bibitem{Gamba_Pelizzari}
A.~Gamba and P.~Pelizzari.
\newblock Utility based pricing of contingent claims in incomplete markets.
\newblock {\em Applied Math. Finance}, 9:241--260, 2002.

\bibitem{Geanakoplos_Polemarchakis}
J.~Geanakoplos and H.~Polemarchakis.
\newblock Existence, regularity and constrained suboptimality of competitive
  allocations when the asset market is incomplete.
\newblock In W.~Heller and D.~Starrett, editors, {\em Essays in Honour of K.
  Arrow, Vol. III}. Cambridge University Press (Cambridge, UK), 1986.

\bibitem{Hobson}
D.~Hobson.
\newblock A survey of mathematical finance.
\newblock {\em Proc. R. Soc. Lond. A}, 460:3369--3401, 2004.

\bibitem{Karatzas_Shreve}
I.~Karatzas and S.~Shreve.
\newblock {\em Methods of Mathematical Finance}.
\newblock Springer (New York), 1998.

\bibitem{Linhart}
P.~B. Linhart.
\newblock Bargaining solutions with non-standard objectives.
\newblock {\em Rev. Econ. Design}, 6:225--239, 2001.

\bibitem{Magill_Shafer}
M.~Magill and W.~Shafer.
\newblock Incomplete markets.
\newblock In W.~Hildenbrand and H.~Sonnenschein, editors, {\em Handbook of
  Mathematical Economics, Vol. IV}. North-Holland (Amsterdam), 1991.

\bibitem{Mao}
X.~Mao.
\newblock {\em Stochastic differential equations and applications}.
\newblock Ellis Horwood (Chichester, UK), 1997.

\bibitem{MasCollel_Whinston_Whinston}
A.~Mas-Colell, M.~Whinston, and J.~R. Green.
\newblock {\em Microeconomic Theory}.
\newblock Oxford University Press (USA), 1995.

\bibitem{Staum}
J.~Staum.
\newblock Incomplete markets.
\newblock In J.~R. Birge and V.~Linetsky, editors, {\em Financial Engineering},
  volume~15 of {\em Handbooks in Operations Research and Management Science}.
  Elsevier (Amsterdam), 2008.

\bibitem{Xanthopoulos_Yannacopoulos}
S.~Z. Xanthopoulos and A.~N. Yannacopoulos.
\newblock Scenarios for price determination in incomplete markets.
\newblock {\em International Journal of Theoretical and Applied Finance},
  11:415--445, 2008.

\end{thebibliography}
\bibliographystyle{plain} 

\end{document}